\PassOptionsToPackage{unicode}{hyperref}
\PassOptionsToPackage{hyphens}{url}
\PassOptionsToPackage{dvipsnames,svgnames,x11names}{xcolor}
\documentclass[
]{article}

\usepackage{amsmath,amssymb}
\usepackage{iftex}
\ifPDFTeX
  \usepackage[T1]{fontenc}
  \usepackage[utf8]{inputenc}
  \usepackage{textcomp} % provide euro and other symbols
\else % if luatex or xetex
  \usepackage{unicode-math}
  \defaultfontfeatures{Scale=MatchLowercase}
  \defaultfontfeatures[\rmfamily]{Ligatures=TeX,Scale=1}
\fi
\usepackage{lmodern}
\ifPDFTeX\else  
\fi
\IfFileExists{upquote.sty}{\usepackage{upquote}}{}
\IfFileExists{microtype.sty}{% use microtype if available
  \usepackage[]{microtype}
  \UseMicrotypeSet[protrusion]{basicmath} % disable protrusion for tt fonts
}{}
\makeatletter
\@ifundefined{KOMAClassName}{% if non-KOMA class
  \IfFileExists{parskip.sty}{%
    \usepackage{parskip}
  }{% else
    \setlength{\parindent}{0pt}
    \setlength{\parskip}{6pt plus 2pt minus 1pt}}
}{% if KOMA class
  \KOMAoptions{parskip=half}}
\makeatother
\usepackage{xcolor}
\makeatletter
\ifx\paragraph\undefined\else
  \let\oldparagraph\paragraph
  \renewcommand{\paragraph}{
    \@ifstar
      \xxxParagraphStar
      \xxxParagraphNoStar
  }
  \newcommand{\xxxParagraphStar}[1]{\oldparagraph*{#1}\mbox{}}
  \newcommand{\xxxParagraphNoStar}[1]{\oldparagraph{#1}\mbox{}}
\fi
\ifx\subparagraph\undefined\else
  \let\oldsubparagraph\subparagraph
  \renewcommand{\subparagraph}{
    \@ifstar
      \xxxSubParagraphStar
      \xxxSubParagraphNoStar
  }
  \newcommand{\xxxSubParagraphStar}[1]{\oldsubparagraph*{#1}\mbox{}}
  \newcommand{\xxxSubParagraphNoStar}[1]{\oldsubparagraph{#1}\mbox{}}
\fi
\makeatother

\usepackage{longtable,booktabs,array}
\usepackage{calc} % for calculating minipage widths
\usepackage{etoolbox}
\makeatletter
\patchcmd\longtable{\par}{\if@noskipsec\mbox{}\fi\par}{}{}
\makeatother
\IfFileExists{footnotehyper.sty}{\usepackage{footnotehyper}}{\usepackage{footnote}}
\makesavenoteenv{longtable}
\usepackage{graphicx}
\makeatletter
\def\maxwidth{\ifdim\Gin@nat@width>\linewidth\linewidth\else\Gin@nat@width\fi}
\def\maxheight{\ifdim\Gin@nat@height>\textheight\textheight\else\Gin@nat@height\fi}
\makeatother
\setkeys{Gin}{width=\maxwidth,height=\maxheight,keepaspectratio}
\makeatletter
\def\fps@figure{htbp}
\makeatother
\NewDocumentCommand\citeproctext{}{}

\makeatletter
 \let\@cite@ofmt\@firstofone
 \def\@biblabel#1{}
 \def\@cite#1#2{{#1\if@tempswa , #2\fi}}
\makeatother
\newlength{\cslhangindent}
\newlength{\csllabelwidth}
\newenvironment{CSLReferences}[2] % #1 hanging-indent, #2 entry-spacing
 {\begin{list}{}{%
  \setlength{\itemindent}{0pt}
  \setlength{\leftmargin}{0pt}
  \setlength{\parsep}{0pt}
  \ifodd #1
   \setlength{\leftmargin}{\cslhangindent}
   \setlength{\itemindent}{-1\cslhangindent}
  \fi
  \setlength{\itemsep}{#2\baselineskip}}}
 {\end{list}}
\usepackage{calc}

\usepackage{arxiv}
\usepackage{orcidlink}
\usepackage{amsmath}
\usepackage[T1]{fontenc}
\makeatletter
\@ifpackageloaded{caption}{}{\usepackage{caption}}
\AtBeginDocument{%
\ifdefined\contentsname
  \renewcommand*\contentsname{Table of contents}
\else
  \newcommand\contentsname{Table of contents}
\fi
\ifdefined\listfigurename
  \renewcommand*\listfigurename{List of Figures}
\else
  \newcommand\listfigurename{List of Figures}
\fi
\ifdefined\listtablename
  \renewcommand*\listtablename{List of Tables}
\else
  \newcommand\listtablename{List of Tables}
\fi
\ifdefined\figurename
  \renewcommand*\figurename{Figure}
\else
  \newcommand\figurename{Figure}
\fi
\ifdefined\tablename
  \renewcommand*\tablename{Table}
\else
  \newcommand\tablename{Table}
\fi
}
\@ifpackageloaded{float}{}{\usepackage{float}}
\floatstyle{ruled}
\@ifundefined{c@chapter}{\newfloat{codelisting}{h}{lop}}{\newfloat{codelisting}{h}{lop}[chapter]}
\floatname{codelisting}{Listing}

\makeatother
\makeatletter
\@ifpackageloaded{caption}{}{\usepackage{caption}}
\@ifpackageloaded{subcaption}{}{\usepackage{subcaption}}
\makeatother
\makeatletter
\@ifpackageloaded{algorithm}{}{\usepackage{algorithm}}
\makeatother
\makeatletter
\@ifpackageloaded{algpseudocode}{}{\usepackage{algpseudocode}}
\makeatother
\makeatletter
\@ifpackageloaded{caption}{}{\usepackage{caption}}
\makeatother

\ifLuaTeX
  \usepackage{selnolig}  % disable illegal ligatures
\fi
\usepackage{bookmark}

\IfFileExists{xurl.sty}{\usepackage{xurl}}{} % add URL line breaks if available
\hypersetup{
  pdftitle={Advancing Geological Carbon Storage Monitoring with 3D Digital Shadow Technology},
  pdfauthor={Abhinav Prakash Gahlot; Rafael Orozco; Felix J. Herrmann},
  colorlinks=true,
  linkcolor={blue},
  filecolor={Maroon},
  citecolor={Blue},
  urlcolor={Blue},
  pdfcreator={LaTeX via pandoc}}

\newcommand{\runninghead}{A Preprint }
\title{Advancing Geological Carbon Storage Monitoring with 3D Digital
Shadow Technology}
\def\asep{\\\\\\ } % default: all authors on same column
\author{\textbf{Abhinav Prakash Gahlot}\\\\Georgia Institute of
Technology\\\\\asep\textbf{Rafael Orozco}\\\\Georgia Institute of
Technology\\\\\asep\textbf{Felix J. Herrmann}\\\\Georgia Institute of
Technology\\\\}
\date{}
\begin{document}
\maketitle

\floatname{algorithm}{Algorithm}

\newcommand{\argmin}{\mathop{\mathrm{argmin}\,}\limits}
\newcommand{\argmax}{\mathop{\mathrm{argmax}\,}\limits}

\[
\def\textsc#1{\dosc#1\csod} 
\def\dosc#1#2\csod{{\rm #1{\small #2}}} 
\]

\section{Introduction}\label{introduction}

To limit global warming and achieve climate goals, the IPCC highlights
the need for technologies capable of removing gigatonnes of
CO\textsubscript{2} annually (IPCC, Panel on Climate Change) (2018)),
with Geological Carbon Storage (GCS) playing a key role in this effort.
GCS captures CO\textsubscript{2} emissions and stores them in deep
geological formations. As a scalable solution for net-negative emissions
(Ringrose 2020, 2023), GCS is considered essential for reducing
anthropogenic CO\textsubscript{2}. Its safety depends on accurate
monitoring of subsurface CO\textsubscript{2} flow and its long-term
behavior. Advanced time-lapse seismic imaging is particularly crucial
for tracking CO\textsubscript{2} migration, detecting leaks, optimizing
operations, and ensuring the integrity of storage sites.

A Digital Shadow (Gahlot, Orozco, et al. 2024) is a framework for
monitoring CO\textsubscript{2} storage projects. It integrates field
data from sources such as seismic and borehole well measurements to
track changes in CO\textsubscript{2} saturation over time. This
framework updates a digital model of the CO\textsubscript{2} using
machine learning-assisted data assimilation techniques (Spantini,
Baptista, and Marzouk 2022; Gahlot, Li, et al. 2024; Gahlot, Orozco, et
al. 2024), including generative AI and nonlinear ensemble Bayesian
filtering. By employing a Bayesian approach, it addresses uncertainties
in reservoir properties, such as permeability, and incorporates this
lack of information into uncertainties of plume predictions.

Compared to 2D monitoring, 3D monitoring is critical for accurately
assessing and controlling GCS projects, as it provides a detailed,
spatially accurate representation of induced changes in the subsurface.
In GCS projects, CO\textsubscript{2} injection and plume migration occur
in three dimensions, making 3D seismic imaging and reservoir modeling
essential for capturing the full extent of CO\textsubscript{2} behavior.
This abstract extends the uncertainty-aware 2D Digital Shadow framework
(Gahlot, Orozco, et al. 2024) by incorporating 3D seismic imaging and 3D
reservoir modeling to improve decision-making and mitigate risks
associated with subsurface storage.

\section{Methodology}\label{methodology}

To develop an uncertainty-aware 3D digital shadow, we follow the
formulation in Gahlot, Orozco, et al. (2024) for the plume dynamics and
observation models,

\begin{equation}\phantomsection\label{eq-dynamics}{
\begin{aligned}
\mathbf{x}_{k} & = \mathcal{M}_k\bigl(\mathbf{x}_{k-1}, \boldsymbol{\kappa}_k; t_{k-1}, t_k\bigr)\\
               & = \mathcal{M}_k\bigl(\mathbf{x}_{k-1}, \boldsymbol{\kappa}_k\bigr),\ \boldsymbol{\kappa}_k\sim p(\boldsymbol{\kappa}) \quad \text{for}\quad k=1\cdots K.
\end{aligned}
}\end{equation}

where \(\mathbf{x}_k\) is the time-varying spatial distribution for the
CO\textsubscript{2} saturation and pressure perturbation measured at
time instants \(k=1\cdots K\), \(\mathcal{M}_k\) represents multi-phase
fluid-flow simulations that transition the state from the previous
timestep, \(t_{k-1}\), to the next timestep, \(t_k\), for all \(K\)
time-lapse timesteps. The symbol \(\boldsymbol{\kappa}\) represents the
spatial distribution of the permeability, which is highly heterogeneous
(Ringrose 2020, 2023), and is assumed to be known only in terms of its
statistical distribution \(p(\boldsymbol{\kappa})\). To incorporate this
lack of knowledge, every time the dynamics operator is applied new
samples of the subsurface permeability distribution,
\(\boldsymbol{\kappa}\sim p(\boldsymbol{\kappa})\) are sampled while the
CO\textsubscript{2} saturation and reservoir pressure perturbations,
collected at the previous timestep \(k-1\) in the vector,
\(\mathbf{x}_{k-1}\), are advanced to the next timestep,
\(\mathbf{x}_{k}\).

While the dynamics can be accurately modeled through reservoir
simulations, relying solely on such simulations is impractical due to
the inherent stochasticity of the permeability, which remains poorly
constrained. To enhance estimates of CO\textsubscript{2} plume behavior,
real-time monitoring is essential. Time-lapse data from 3D surface
seismic measurements (Lumley 2010) provide critical information, which
we incorporate into our characterization of the CO\textsubscript{2}
plumes. These time-lapse datasets are integrated within a framework
designed to better capture the state of CO\textsubscript{2} plume,
accounting for the uncertainties in the subsurface reservoir properties.
The seismic datasets are modeled as

\begin{equation}\phantomsection\label{eq-obs}{
\mathbf{y}_k = \mathcal{H}_k(\mathbf{x}_k) + \boldsymbol{\epsilon}_k , \ \boldsymbol{\epsilon}_k \sim p(\boldsymbol{\epsilon}) \quad\text{for}\quad k=1,2,\cdots,K
}\end{equation}

where \(\mathbf{y}_k\) represents the 3D seismic data at timestep \(k\)
and \(\boldsymbol{\epsilon}_k\) is colored Gaussian noise added to the
seismic shot records prior to reverse-time migration. To assimilate the
fluid-flow simulations and the time-lapse seismic images, we use a
nonlinear Bayesian filtering framework, using neural posterior density
estimation to approximate the posterior distribution of the
CO\textsubscript{2} plume state conditioned on the time-lapse seismic
data. This approach leverages advanced techniques from sequential
Bayesian inference and amortized neural posterior density estimation
(Papamakarios et al. 2021) in a principled manner. To estimate the
posterior distribution of the CO\textsubscript{2} plume given seismic
observations, we use conditional Normalizing Flows (CNFs) (Gahlot,
Orozco, et al. 2024). In this simulation-based framework, we carry out
simulations of the state dynamics and observations (cf.~equations
\ref{eq-dynamics} and \ref{eq-obs}) and train the CNFs on these pairs.

\section{Synthetic case study}\label{synthetic-case-study}

To test our method, we use a synthetic 3D Earth model from the Compass
model (E. Jones et al. 2012) representative for the North Sea. We select
a subset of the Compass model that includes subsurface structures
suitable for CO\textsubscript{2} injection. The studied subset spans a
physical size of 3.2 km³, discretized into a grid of dimensions
\(128 \times 128 \times 128\). To initialize the probabilistic digital
shadow, we require an initial ensemble of potential plume scenarios.
Since the reservoir properties of the CO\textsubscript{2} storage sites
are unknown, we establish a probabilistic baseline for the 3D
permeability properties of the site and use this distribution to create
the plume initialization. This permeability distribution is derived by
first performing probabilistic 3D full-waveform inference (Orozco,
Siahkoohi, et al. 2024) on the site, assuming a baseline seismic survey
was conducted prior to injection. The resulting distribution of the
unknown acoustic velocity models is then converted into \(128\)
permeability samples using the empirical relationship described in
Gahlot, Orozco, et al. (2024).

\subsection{Multi-phase flow
simulations}\label{multi-phase-flow-simulations}

The flow simulations are carried out using an open-source tool,
JutulDarcy
\href{https://github.com/sintefmath/JutulDarcy.jl}{JutulDarcy.jl}
(Møyner, Bruer, and Yin 2023). Initially, the reservoir is filled with
brine, and then supercritical CO\textsubscript{2} is injected at the
rate chosen such that the storage capacity of the reservoir rocks does
not exceed \(3\%\). The injection depth is approximately \(1130 m\). The
simulation is performed over a 4-year interval for each time-step,
\(t_k\). It outputs predicted CO\textsubscript{2} saturation for each of
the \(N=128\) ensemble members. One of the training (predicted)
saturations is shown in Figure 1(c).

\subsection{Seismic simulations}\label{seismic-simulations}

The output from the 128 flow simulations are converted to changes in the
subsurface acoustic properties using the Patchy saturation model
(Gahlot, Orozco, et al. 2024). To monitor these acoustic changes, we
shoot 4D seismic surveys using 16000 receivers and 16 sources with a
dominant frequency of 24 Hz and a recording length of 3 sec.~We added 18
dB SNR of colored Gaussian noise to the shot records. The nonlinear wave
simulations and imaging are performed using the open-source software
package \href{https://github.com/slimgroup/JUDI.jl}{JUDI.jl} (Witte et
al. 2019; Louboutin et al. 2023).

\subsection{CNFs training}\label{cnfs-training}

The \(128\) members of the ensemble serve as training pairs. These pairs
are created from the above simulations and consist of a forecasted plume
and the associated 4D seismic observation. For the CNF training, we
implemented a first-of-its-kind 3D conditional normalizing flow in the
open-source package
\href{https://github.com/slimgroup/InvertibleNetworks.jl}{InvertibleNetworks.jl}
(Orozco, Witte, et al. 2024). Crucially, the scalability of this
generative network to 3D volumes was enabled by the memory-frugal
invertible layers of the normalizing flow. The network weights
\(\boldsymbol{\phi}\) are trained by minimizing the objective
equation~\ref{eq-loss-CNF} with \(\textsc{ADAM}\) (Kingma and Ba 2014)
optimizer and for a total of \(120\) epochs,

\begin{equation}\phantomsection\label{eq-loss-CNF}{
\widehat{\boldsymbol{\phi}} = \mathop{\mathrm{argmin}\,}\limits_{\boldsymbol{\phi}} \frac{1}{M}\sum_{m=1}^M \Biggl(\frac{\Big\|f_{\boldsymbol{\phi}}(\mathbf{x}^{(m)};\mathbf{y}^{(m)})\Big\|_2^2}{2} - \log\Bigl |\det\Bigl(\mathbf{J}^{(m)}_{f_{\boldsymbol{\phi}}}\Bigr)\Bigr |\Biggr).
}\end{equation}

where \(\mathbf{J}\) represents the Jacobian of the network
\(f_{\theta}\) with respect to its input and \(\text M\) is the number
of training samples. For details on the implemenation we refer to
(Gahlot, Orozco, et al. 2024).

\section{Results}\label{results}

\begin{figure}

\begin{minipage}{0.50\linewidth}

\centering{

\includegraphics[width=0.85\textwidth,height=\textheight]{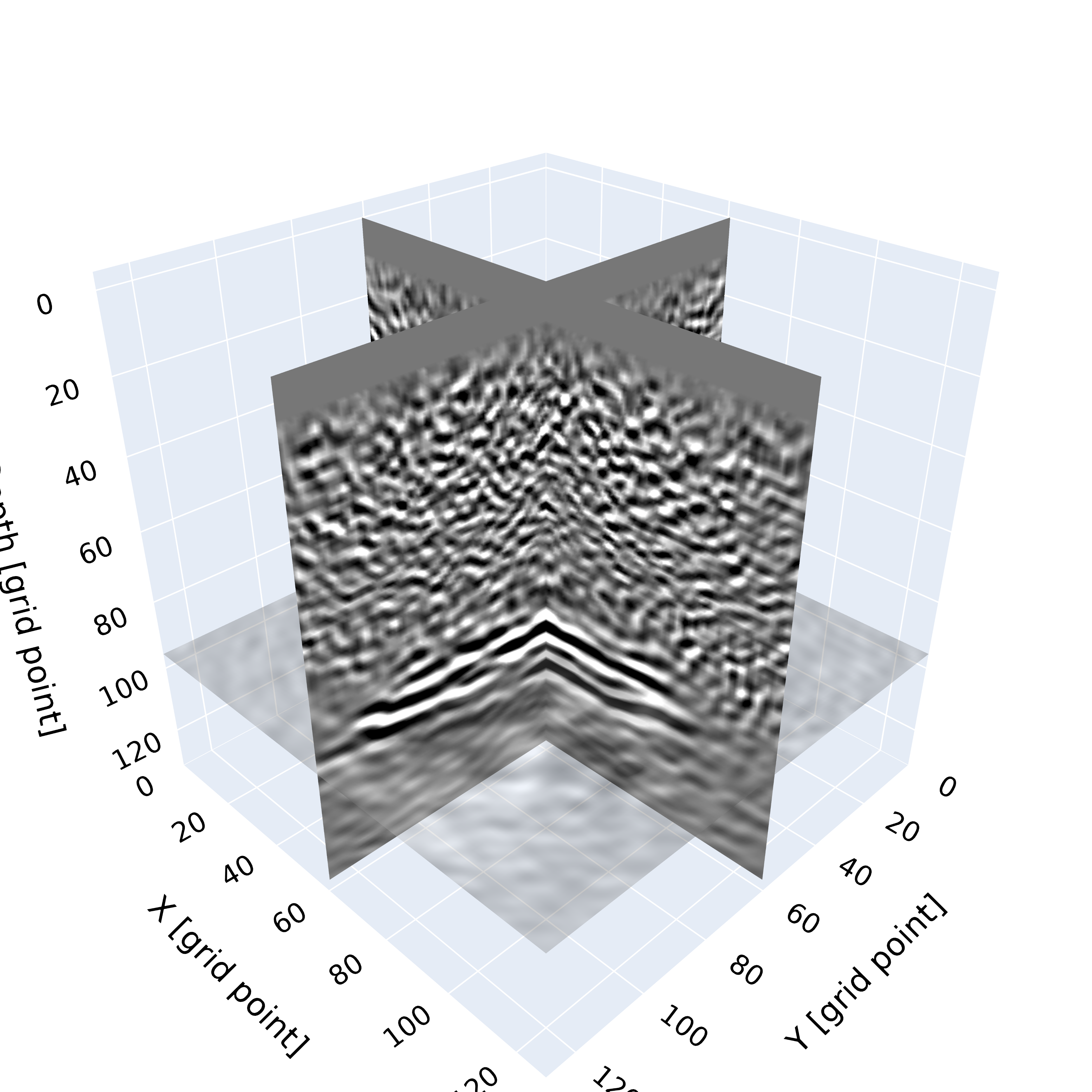}

}

\subcaption{\label{fig-obs}4D seismic observation}

\end{minipage}%
\begin{minipage}{0.50\linewidth}

\centering{

\includegraphics[width=0.85\textwidth,height=\textheight]{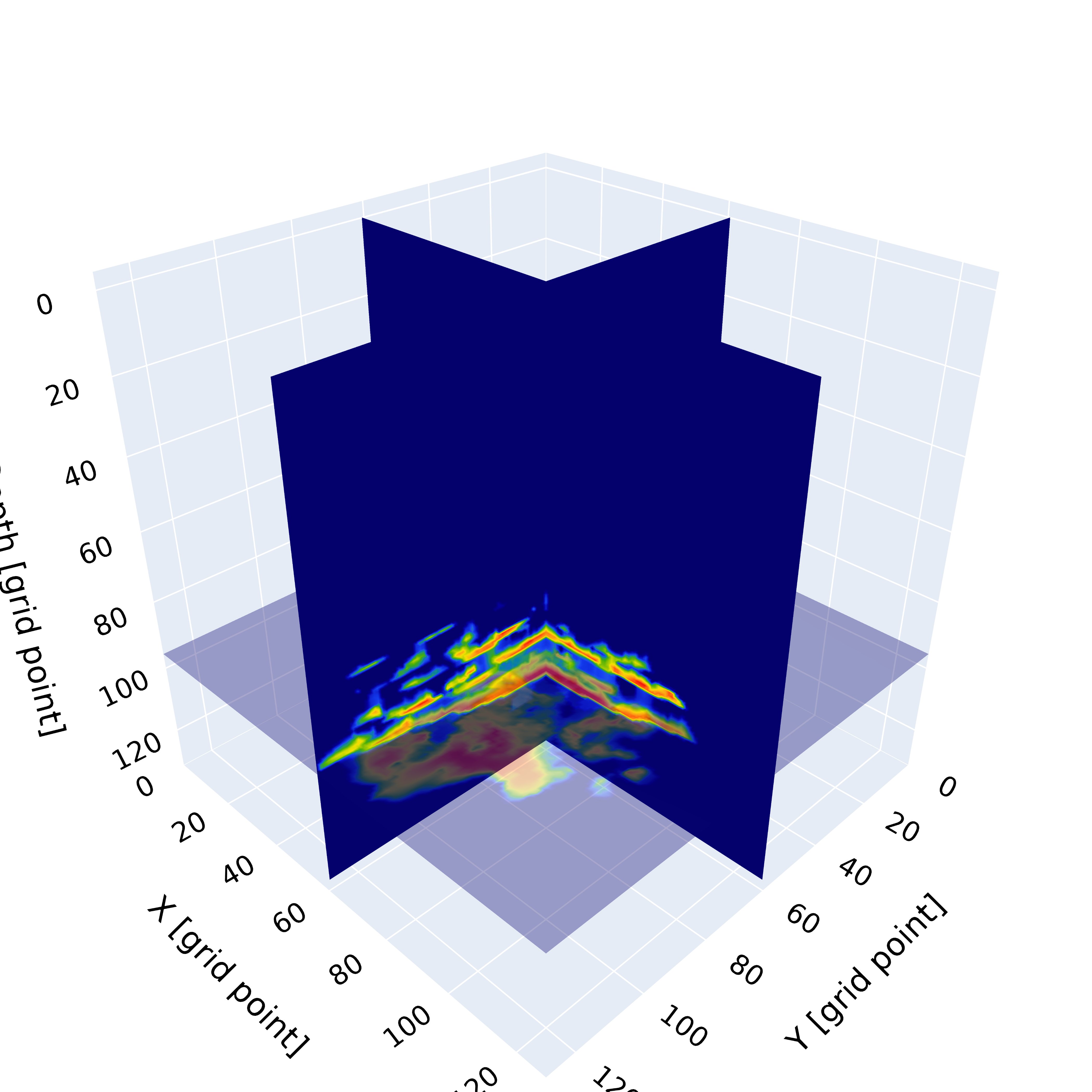}

}

\subcaption{\label{fig-gt}Unknown ground truth plume}

\end{minipage}%
\newline
\begin{minipage}{0.50\linewidth}

\centering{

\includegraphics[width=0.85\textwidth,height=\textheight]{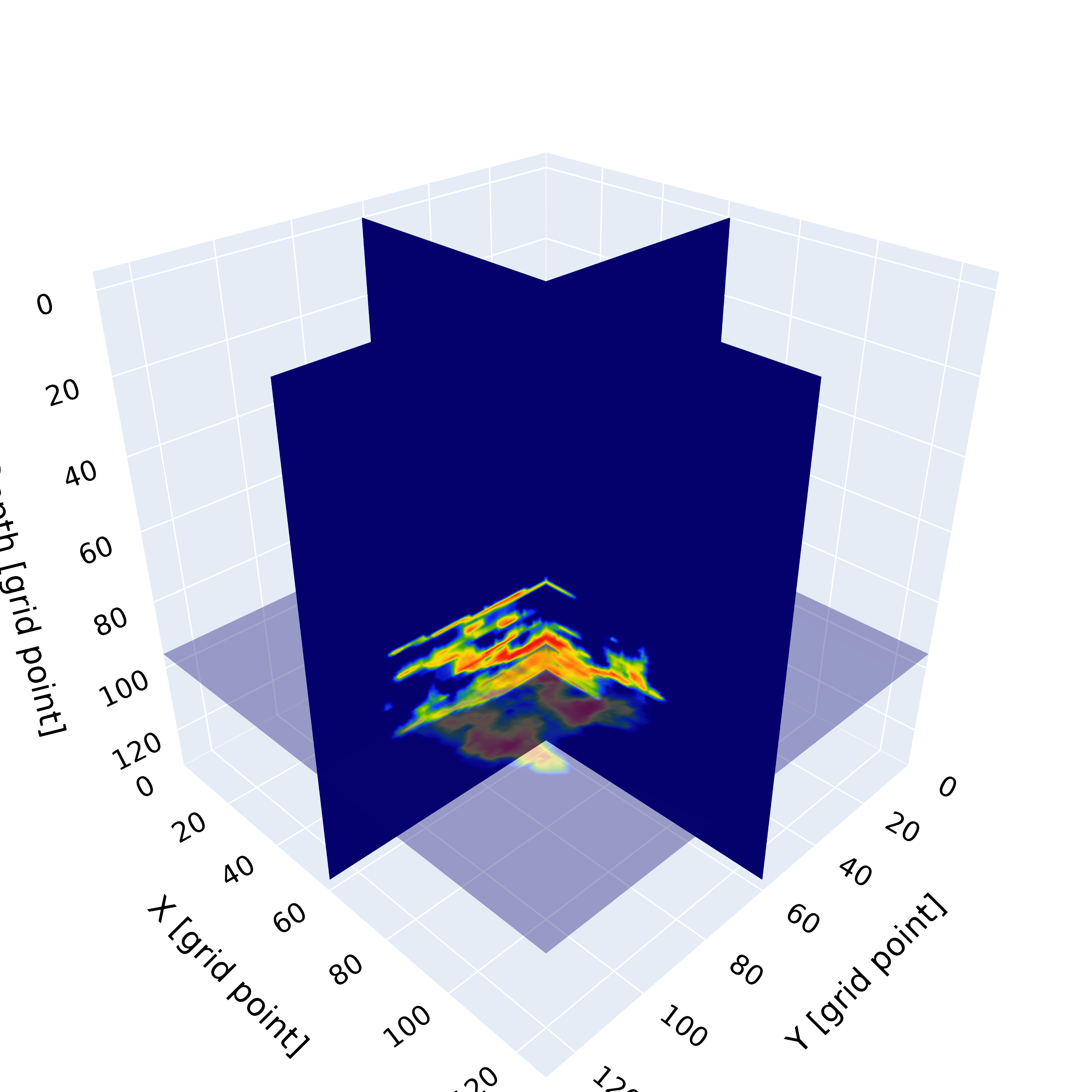}

}

\subcaption{\label{fig-train}Training plume}

\end{minipage}%
\begin{minipage}{0.50\linewidth}

\centering{

\includegraphics[width=0.85\textwidth,height=\textheight]{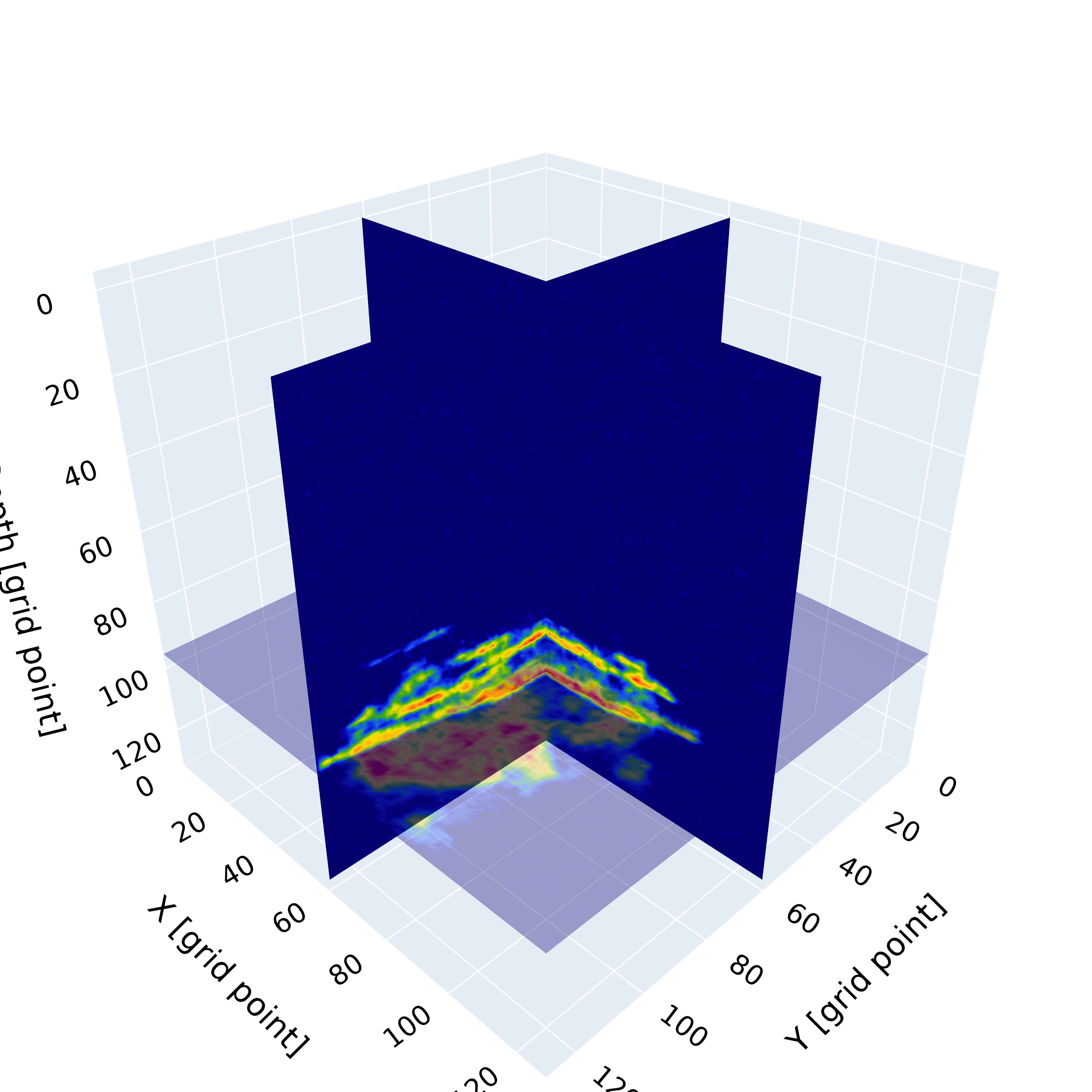}

}

\subcaption{\label{fig-posterior}Assimilated plume}

\end{minipage}%
\newline
\begin{minipage}{0.50\linewidth}

\centering{

\includegraphics[width=0.85\textwidth,height=\textheight]{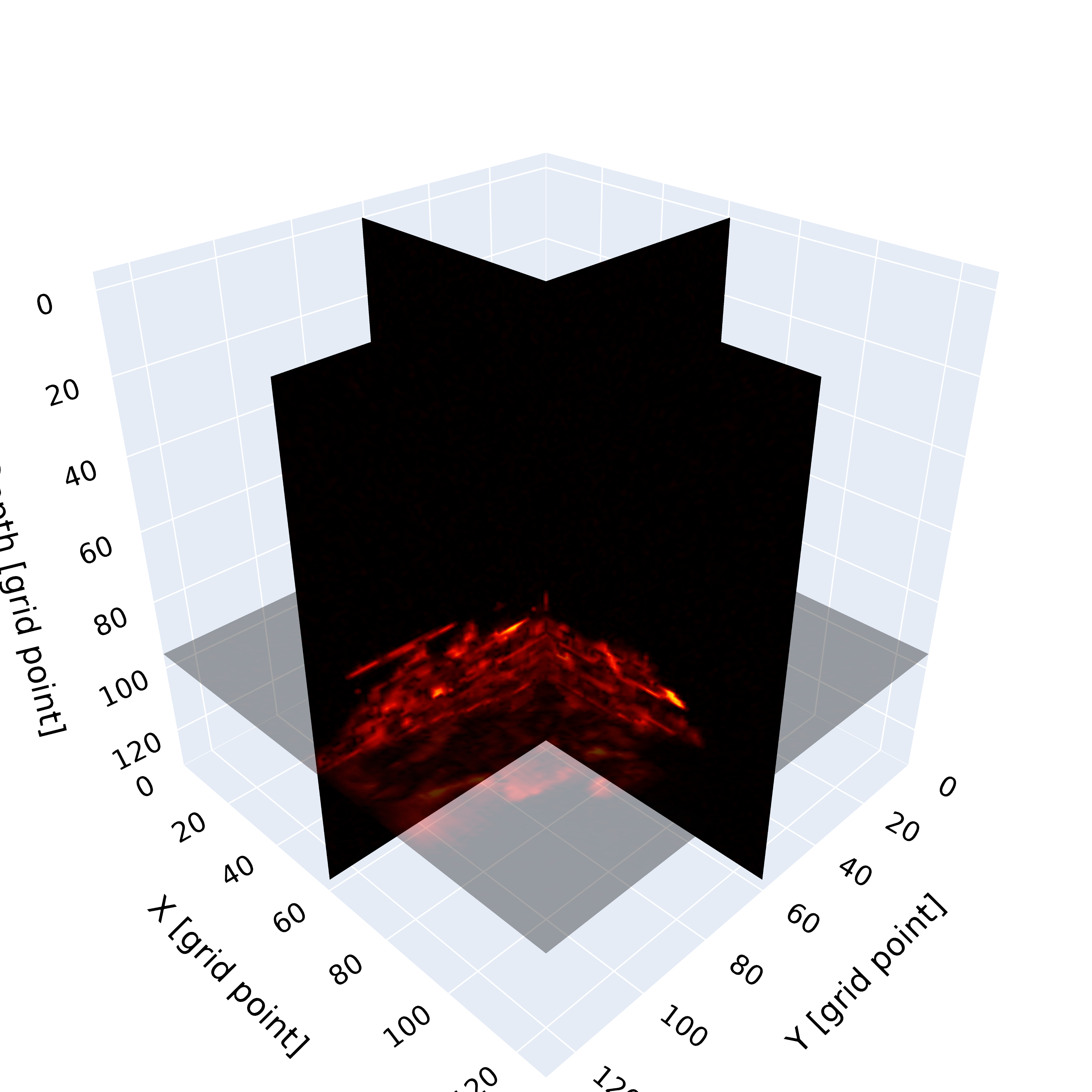}

}

\subcaption{\label{fig-error}Estimated plume error}

\end{minipage}%
\begin{minipage}{0.50\linewidth}

\centering{

\includegraphics[width=0.85\textwidth,height=\textheight]{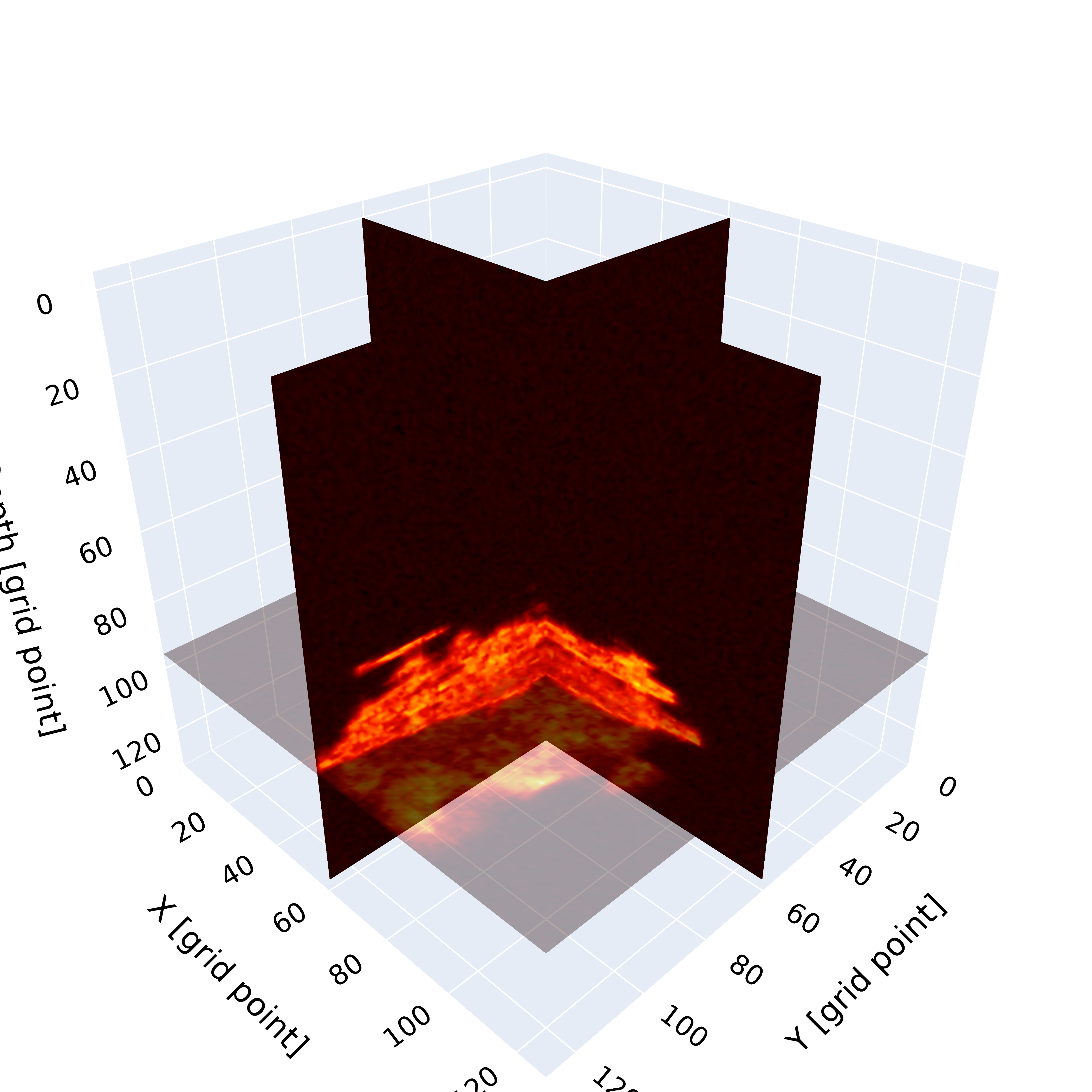}

}

\subcaption{\label{fig-uq}Estimated plume uncertainty}

\end{minipage}%

\caption{\label{fig-sup}\emph{(a)} Time-lapse seismic measurements used
as conditioning input to the trained conditional Normalizing Flow.
\emph{(b)} The ground-truth unknown plume that is the target to monitor
\emph{(c) An ensemble member for the state of the plume used for
training. }(d)* A posterior sample conditioned on the seismic
observation \emph{(e)} Error between the inferred posterior mean and the
ground truth \emph{(f)} The variance between posterior samples.}

\end{figure}

We illustrate the performance of our method on an unseen (not part of
the training) plume shown in figure~\ref{fig-gt}. To calculate posterior
statistics, we use 128 posterior samples to calculate the mean and
standard deviation. figure~\ref{fig-posterior} shows the inferred plume
conditioned on the observed seismic data in figure~\ref{fig-obs}.
figure~\ref{fig-error} shows the error between the mean of the posterior
and the ground-truth plume in figure~\ref{fig-gt}. figure~\ref{fig-uq}
shows the uncertainty of the inferred plume. We highlight that there is
a good correlation between the error and uncertainty, which is an
indication that the uncertainty is well calibrated.

\section{Conclusions}\label{conclusions}

We present a 3D Digital Shadow to monitor a 3D time-lapse
CO\textsubscript{2} flow using 4D seismic data. Our method introduces a
novel implementation of scalable normalizing flows, which facilitates
the first known application of generative modeling to realistically
sized 3D volumes. The results show significant improvements in
CO\textsubscript{2} plume estimates when seismic 4D data was
incorporated at the monitoring step, particularly in bringing the
predicted plume to the correct shape and size shown in the seismic and
ground truth plume. The framework also demonstrates well-calibrated
uncertainty quantification, with strong correlations between inferred
uncertainty and ground-truth errors. This study confirms that the
Digital Shadow, integrating 3D seismic data, effectively tracks
CO\textsubscript{2} plume dynamics and lays the groundwork for a 3D
Digital Twin to optimize CO\textsubscript{2} storage operations.

\section{Acknowledgement}\label{acknowledgement}

This research was carried out with the support of Georgia Research
Alliance, partners of the ML4Seismic Center and in part by the US
National Science Foundation grant OAC 2203821.

\clearpage
\section*{References}\label{references}
\addcontentsline{toc}{section}{References}

\phantomsection\label{refs}

\end{document}